\documentclass[12pt]{iopart}
\usepackage{iopams}  
\usepackage{graphicx}
\usepackage{cite}
\usepackage{mciteplus}
\begin{document}

\title[Uncertainty propagation within the UNEDF models]{Uncertainty propagation within the UNEDF models}

\author{T. Haverinen$^{1,2}$, M. Kortelainen$^{2,1}$}

\address{${}^1$ Helsinki Institute of Physics, P.O. Box 64, FI-00014 University of Helsinki, Finland}
\address{${}^2$ Department of Physics, P.O. Box 35 (YFL), University of Jyvaskyla, FI-40014 Jyvaskyla, Finland}
\eads{\mailto{tiia.k.haverinen@student.jyu.fi}, \mailto{markus.kortelainen@jyu.fi}}

\begin{abstract}
The parameters of the nuclear energy density have to be adjusted to experimental data. 
As a result they carry certain uncertainty which then propagates to calculated
values of observables. In the present work we quantify the statistical uncertainties 
of binding energies, proton quadrupole moments, and proton matter radius
for three UNEDF Skyrme energy density functionals by taking advantage of the knowledge 
of the model parameter uncertainties. We find that the uncertainty of UNEDF models increases 
rapidly when going towards proton or neutron rich nuclei.
We also investigate the impact of each model parameter on the total error budget.


\end{abstract}
\vspace{2pc}

\vspace{2pc}

%\noindent{(Some figures may appear in colour only in the online journal)}

%\submitto{\JPG}
%\maketitle

\section{Introduction}

Among numerous different nuclear many-body models, the nuclear density functional theory 
(DFT)~\cite{(Ben03)} is the only one, which can describe nuclear properties 
microscopically throughout the entire nuclear landscape~\cite{(Erl12)}. 
The cornerstone of the nuclear DFT is the energy density functional (EDF), which incorporates
nucleonic interactions and many-body correlations into a functional constructed from one-body densities and 
currents. The Skyrme EDF, for its part, relies on local nuclear densities and currents, together 
with a set of coupling constants as model parameters. Due to the lack of suitable \emph{ab-initio} 
methods to compute these coupling constants, they must be determined through adjustment to 
experimental data, such as nuclear binding energies and radii. 

During the last couple of decades, numerous Skyrme parameterizations have been obtained from various 
adjustment schemes, see e.g. the list in~\cite{(Dut12)}.
The standard Skyrme EDF has proven to be quite successful, but its limitations have also become 
apparent. For the sake of better accuracy, more reliable predictive power, and for a spectroscopic-quality level, 
one has to move beyond the the standard Skyrme EDF~\cite{(Car08), (Kor08)}. 
Nevertheless, by studying the performance and predictive power of the present EDFs, valuable 
information can be obtained which can be used and applied in the work towards forthcoming novel EDFs~\cite{(Naz16)}.

As with every model parameters optimization procedure, one of the main challenges is to find the best set of
input observables, in order to constrain the parameter space of the model. The predictive power of the EDF highly 
depends on the input data. Therefore, a comprehensive analysis of the impact of input observables
on the parameter space and on the model predictions provides valuable information.
Naturally, the nuclear bulk properties are crucial for general constraining and the data especially 
relating to odd-mass nuclei is important for spectroscopical properties~\cite{(Sch10)}. 
Binding energies, surface thickness, charge radii, single particle energies and energies of 
giant resonances are essential properties of nuclei, and used in various EDF optimization schemes.

All model predictions contain several sources of uncertainties. Roughly speaking, these can be divided into 
two main categories, the systematic model uncertainties and the statistical model uncertainties.
The systematic model uncertainty stems from sources like the model deficiency and input data bias. 
The statistical uncertainty results from the model parameter optimization process.

Despite the importance of uncertainty analysis, error estimate is a rather novel topic 
in low-energy nuclear physics~\cite{(Dob14)}. 
During the last few years, efforts have been made to improve this situation in the EDF 
calculations~\cite{(Gao13),(Sch15), (Rio15), (Nik15), 
(Kor15)}, 
as well as in the domain of \emph{ab-initio} calculations~\cite{(Eks15), (Lah15)}.
Various statistical tools have been applied from traditional methods to more modern ones 
(e.g. the Bayesian framework~\cite{(McD15), (Hig15), (Gra15),(Wes16)}). 
Apart from the fact that uncertainty quantification is an important topic in itself, 
with the help of statistical analysis, information about shortcomings of theoretical 
models and optimization procedures is also obtained. 

In this work, we present the quantitative results for statistical uncertainty propagation 
for three existing UNEDF Skyrme EDF models: the UNEDF0~\cite{(Kor10)}, the UNEDF1~\cite{(Kor12)},
and the UNEDF2~\cite{(Kor14)}.
In particular, we quantify contributions from the model parameters to the total error budget of binding energy 
in isotopic and isotonic chains of nuclei. By analyzing the obtained information we may recognize potential 
frailties of these models. In the present study, two-neutron separation energies are also considered,
as well as proton quadrupole moments and proton matter radii, and the related uncertainties are worked out.
In addition to even-even nuclei, uncertainties related to odd-even nuclei are studied.

This paper is organized as follows. In Sec.~\ref{sec:theorframe} we briefly review the theoretical 
framework related to the topic: Namely, the Skyrme energy density functional and the error propagation. 
In Sec.~\ref{sec:results} we present our results and, finally, conclusions and future perspectives 
are given in Sec.~\ref{sec:concl}.

\section{Theoretical framework}
\label{sec:theorframe}

\subsection{Skyrme energy density functional}
The UNEDF models are based on the Skyrme energy density. The ground state of a nucleus is 
determined in the framework of the Hartree-Fock-Bogoliubov (HFB) theory~\cite{(Ben03),(Rin04)}.
The three parameterizations considered in this work, UNEDF0, UNEDF1 and UNEDF2, were adjusted on a experimental 
data consisting of binding energies for deformed and spherical nuclei, odd-even mass differences, and charge radii.
In addition, latter parameterizations include data on
fission isomer excitation energies and single-particle energies. 

The Skyrme EDF has a form of local energy density functional, stemming from Skyrme energy density. 
It can be written as
\begin{eqnarray}
E  &=& \int  \mathrm{d}^3\textbf{r} \mathcal{H} (\mathbf{r}) \\
\label{eq:SkyrmeE}
&=& \int \mathrm{d}^3\textbf{r} \left[ \mathcal{E}^{\rm kin} (\mathbf{r}) + \chi_0 (\mathbf{r}) + \chi_1 (\mathbf{r}) + \widetilde{\chi} (\mathbf{r}) + \mathcal{E}^{\rm Coul} (\mathbf{r}) \right] \,.
\end{eqnarray}
where the energy density \(\mathcal{H} (\mathbf{r})\) is a time-even, scalar, isoscalar and real function of local densities and their derivatives. 
In the equation~(\ref{eq:SkyrmeE}), the Skyrme energy density has been split into kinetic 
term \(\mathcal{E}^{\rm kin} (\mathbf{r})\), isoscalar (\(t=0\)) and isovector (\(t=1\)) 
particle-hole Skyrme energy densities \(\chi_t (\mathbf{r})\), pairing energy density 
\(\widetilde{\chi} (\mathbf{r})\) and Coulomb term \(\mathcal{E}^{\rm Coul} (\mathbf{r})\). 
The time-even part of the isoscalar and isovector particle-hole Skyrme energy densities is given by
\begin{eqnarray}
\chi_t(\textbf{r}) &=& C_t^{\rho^2 } \rho_t^2 + C_t^{\rho \tau} \rho_t \tau_t 
+ C_t^{J J} \sum_{\mu \nu}\mathbf{J}_{\mu\nu,t} \mathbf{J}_{\mu \nu,t} 
+ C_t^{\rho \Delta \rho} \rho_t \Delta \rho_t \nonumber \\
&+& C_t^{\rho \nabla J} \rho_t \nabla \cdot \mathbf{J}_t \,.
\label{eq:phterm}
\end{eqnarray} 
In the equation~(\ref{eq:phterm}), \(\tau_t\) is the isoscalar or isovector kinetic density 
and \(\mathbf{J}_{\mu\nu,t}\) is the spin-current density tensor. 
Definitions of these densities can be found in reference~\cite{(Ben03)}. 
With the UNEDF models, only the time-even part of the total
energy density was defined and time-odd part of the energy density was set to zero. The energy density is always time-even, also the part called ''time-odd'' - the time-odd energy density means that this part of the energy density is built by using time-odd densities.

The pairing energy density \(\widetilde{\chi} (\mathbf{r})\) used here has the form of
\begin{equation}
\widetilde{\chi} (\textbf{r}) = \frac{1}{4} \sum_{q={\rm n,p}} V_0^{q} 
\Bigl[ 1- \frac{1}{2} \frac{\rho_0 (\mathbf{r})}{\rho_c} \Bigr] \widetilde{\rho}_q^2 (\mathbf{r}) \,,
\end{equation}
where \(V_0^{q}\) (\(q={\rm n,p}\)) are the pairing strength parameters for neutrons and protons,
respectively, and \(\rho_{\rm c}\) was set to the equilibrium density \(0.16 \)\,fm\(^{-3}\). 
All the coefficients \(C_t^{x}\) and \(V_0^{q}\) are real constants, except the coefficients \(C_t^{\rho \rho}\) 
which depend on the isoscalar density so that
\begin{equation}
C_t^{\rho \rho} = C_{t0}^{\rho \rho} + C_{t{\rm D}}^{\rho \rho} \rho_0^{\gamma} \,.
\end{equation}
Altogether, there are 13 independent constants from Skyrme energy density and two constants from pairing, namely 
\begin{equation}
\label{eq:parameters}
 \{ C_{t0}^{\rho \rho}, C_{t\rm{D}}^{\rho \rho}, 
C_{t}^{\rho \Delta \rho}, C_t^{\rho \tau}, C_t^{J^2}, C_{t}^{\rho \nabla J} \}_{t=0,1} \, , \, \gamma , V_0^{\rm n} \, $ and $ V_0^{\rm p} \,. 
\end{equation}
Seven of these parameters for \(t=0\) and \(t=1\), \(C_{t0}^{\rho \rho}\), \(C_{t{\rm D}}^{\rho \rho}\), \(C_t^{\rho \tau}\) 
and \(\gamma\), can be written with the help of the infinite nuclear matter 
parameters~\cite{(Kor10), (Cha97)}. 
All in all, the model depends on 15 independent parameters, namely 
\begin{eqnarray}
\label{eq:parameters2}
 &\rho_{\rm c}, \frac{E^{\rm NM}}{A}, K^{\rm NM}, a_{\rm sym}^{\rm NM}, L_{\rm sym}^{\rm NM}, M_{s}^{*}, 
M_v^*, C_{0}^{\rho \Delta \rho}, \nonumber \\
& C_{1}^{\rho \Delta \rho}, V_0^{\rm n}, V_{0}^{\rm p}, C_{0}^{\rho \nabla J}, 
C_{1}^{\rho \nabla J}, C_{0}^{JJ} \textrm{, and } C_{1}^{JJ} \,,
\end{eqnarray} 
which were optimized in the previous works, in 
references~\cite{(Kor10), (Kor12), (Kor14)}. 
Here \(\rho_c\) is the saturation density, \(E^{\rm NM}/A\) represents the total 
energy per nucleon at equilibrium, \(K^{\rm NM}\) is the nuclear matter incompressibility, 
\(a_{\rm sym}^{\rm NM}\) is the symmetry energy coefficient, 
\(L_{\rm sym}^{\rm NM}\) describes the slope of the symmetry energy, 
\(M_s^*\) is the isoscalar effective mass and the last one, \(M_v^*\), is the isovector effective mass.

In the present work, we have compared calculated theoretical binding energies 
to the experimental ones from~\cite{(Aud12)}. In order to obtain experimental 
nuclear binding energies, the experimental atomic masses were corrected 
by taking into account the electron binding energies, approximated as
\begin{equation}
\label{eq:B_E}
B_{\rm E} \approx -1.433  \times 10^{-5} Z^{2.39} \,{\rm MeV}\,.
\end{equation}

\subsection{Propagation of error}

\begin{table}
\caption[UNEDF parameters]{The parameters of UNEDF models  used in the sensitivity analysis.
Here "x" indicates  parameter was included in sensitivity analysis. 
Parameters which were fixed during the whole optimization procedure are denoted as
"--", and the rest of the parameters are those which hit the boundary values during optimization.
The index \(t=0,1\) separates the isoscalar and isovector terms.}
\centering
\begin{tabular}{lccccccccccc}
\br 
EDF & \(\rho_c \) & \(\frac{E^{\rm NM}}{A} \) & \(K^{\rm NM}\) & \(a_{\rm sym}^{\rm NM}\) 
& \(L_{\rm sym}^{\rm NM}\) & \( 1/M_s^{*}  \) & \(1/M_v^*\) &  \(C_{t}^{\rho \Delta \rho} \) 
&  \(V_0^{\rm n,p}\) & \(C_{t}^{\rho \nabla J}\) &  \(C_{t}^{JJ}\)  \\
\mr
UNEDF0 & x & x &   & x & x &   & -- & x & x & x  & --  \\
UNEDF1 & x &   &   & x & x & x & -- & x & x & x  & --  \\
UNEDF2 & x &   & x & x &   & x & -- & x & x & x  & x \\
\br
\end{tabular}
\label{tab:unedfpara}
\end{table}

The UNEDF parameterizations
were accompanied by sensitivity analysis, providing covariance matrix of the model parameters. 
This allows to calculate the standard deviation of any observable predicted by the model.
In the present work we consider the statistical errors on binding energies and on two-neutron separation energies.

The statistical standard deviation \(\sigma\) of an observed variable \(y\) is given by
\begin{equation}
\label{eq:sigma2}
\sigma^2(y) = \sum_{i,j=1}^n $Cov$ (x_i, x_j) \left[ \frac{\partial y}{\partial x_i}\right] \left[\frac{\partial y}{\partial x_j}\right] \,,
\end{equation}
where \($Cov$ (x_i, x_j)\) is the covariance matrix element between the model parameters \(x_i\) and \(x_j\), 
and $n$ is the number of model parameters. The covariance matrix \($Cov$ (x_i, x_j)\) is related to the 
corresponding correlation matrix \($Corr$ (x_i, x_j)\) as
\begin{equation}
\label{eq:covcorr}
$Cov$ (x_i, x_j) = $Corr$ (x_i, x_j) \sigma_{x_i}\sigma_{x_j} \,,
\end{equation}
where \(\sigma_{x_i}\) and \(\sigma_{x_j}\) are the standard deviations of parameters \(x_i\) and \(x_j\), 
respectively.
The correlation matrices of the UNEDF models and the  standard deviations of the model 
parameters are given in \ref{App:AppendixA}. 

The standard deviation in equation~(\ref{eq:sigma2}) contains a sum of terms connected to the
model parameters. Due to correlations between model parameters, off-diagonal components
has to be also taken into account. By diagonalizing the covariance matrix (or, equivalently,
the curvature matrix of $\chi^2({\bf x})$ function) it is possible analyze eigenmodes,
as was demonstrated in reference~\cite{(Fat11)}. This method was also used in analysis
of DD-PC1 functional uncertainties~\cite{(Nik15)}.
With application of an orthogonal transformation, which diagonalizes
the covariance matrix, one obtains the square of the standard deviation \(\sigma^2(y)\)
expressed as a sum over eigenvalues multiplied by corresponding eigenvectors and 
partial derivatives of $y$. 

Some of the UNEDF model parameters have been excluded from the sensitivity analysis.
Table~\ref{tab:unedfpara} lists those UNEDF parameters which were included in the sensitivity 
analysis~\cite{(Kor10), (Kor12), (Kor14)} (marked with \(x\)), 
those which were fixed during the whole optimization procedure (-) and those which ended 
up at their boundaries during the optimization (empty space). 
Sensitivity analysis can not be performed for fixed parameters or those which drifted onto 
the boundary during the optimization. 
However, those parameters which were included in sensitivity analysis have a visible contribution to the statistical
error of an observable and
their contribution to the total error budget can be calculated from equation~(\ref{eq:sigma2}).

\subsection{Numerical methods}
In the present work we used the code HFBTHO~\cite{(Sto05), (Sto13)} to calculate
observables and their statistical errors. The program solves the Hartree-Fock-Bogoliubov equations for
Skyrme EDFs in the axially symmetric harmonic oscillator basis. Time-reversal and parity symmetries 
were assumed. Because particle number is not a good quantum number in HFB theory, we used the Lipkin-Nogami 
method to restore it approximately. The HFB equations were solved in a basis consisting of 20 oscillator 
shells and the convergence criteria was set to \(10^{-7}\). This means that the desired accuracy has been 
reached when the norm of the HFB matrix difference between two consecutive iterations is less than \(10^{-7}\). 
Both of the Coulomb terms, \(\mathcal{E}^{Coul}_{dir}\) and \(\mathcal{E}^{Coul}_{exchange}\) were used, 
but the exchange term was calculated by using Slater approximation. A rough position of the energy 
minimum, with respect of quadrupole deformation, was first located from a constrained HFB calculation. 
Then, an unconstrained HFB calculation was performed in order to converge to the precise position of 
the energy minimum.

In order to obtain standard errors one has to calculate the derivatives of an observable \(y({\bf x})\) 
with respect to the model parameters \(x_i\), as in equation~(\ref{eq:sigma2}). 
In the present work these derivatives were approximated by a finite differences. That is
\begin{equation}
\label{eq:finitediff}
\frac{\partial y}{\partial x_i} \approx \frac{y(x_1,x_2,\ldots,x_i
+\Delta x_i,\ldots,x_n)-y({\bf x})}{\Delta x_i} \,,
\end{equation}
where value of $i$th parameter has been shifted by amount of $\Delta x_i$ from the model base values.
The rounded values of UNEDF parameters and corresponding shifts $\Delta x_i$ have been listed on
Table~\ref{tab:parsteps}. We tested that the computed statistical errors remained essentially the same
when shift parameters $\Delta x_i$ were slightly varied.

Lastly, we recall that standard deviation does not measure the total uncertainty of a model. 
Another main ingredient, namely the systematic error, is much more challenging to assess. It
can be addressed e.g. by studying a dispersion of different predictions given by various Skyrme EDF models~\cite{(Erl12),(Kor13)}. However, due to lack of exact reference model, precise
systematic errors are not within one's reach.

\section{Results}
\label{sec:results}

\subsection{Binding energy residuals}

\begin{figure}[htb]
\centering
\includegraphics[width=0.8\textwidth]{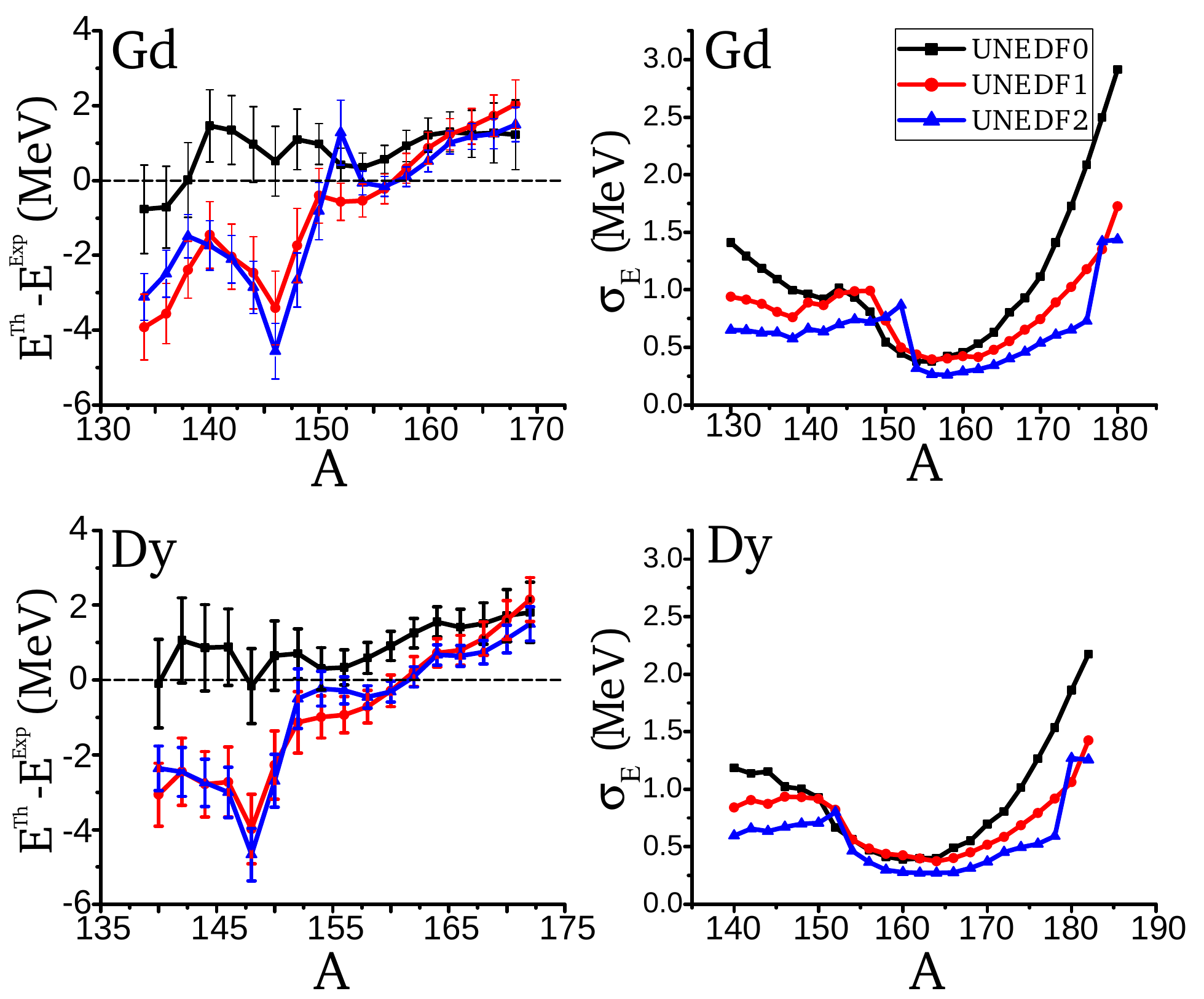}
\caption[The residuals for Gd ad Dy]{(Color online) Differences between theoretical 
and experimental binding energies for even-even dysprosium and gadolinium isotopes with
error bars representing statistical model error
(on the left) and related statistical model error (on the right) as a function of the mass number.}
\label{fig:gddy_residuals}
\end{figure}

The differences between the theoretical and experimental binding energies for even-even gadolinium and
dysprosium isotopes are shown in figure~\ref{fig:gddy_residuals}. The error bars represent
the calculated theoretical standard deviations, and they are also given as a function of the mass number 
in the graphs on the right. The uncertainties are given for all the calculated binding energies,
including also those nuclei for which the experimental binding energy is not currently known. 
As it can be
seen, UNEDF0 gives more consistent results with the measured experimental energies for lighter isotopes, whereas 
UNEDF1 and UNEDF2 seem to improve their accuracy considerably for heavier isotopes. Most of the 
theoretical results do not overlap with the experimental values, even when including error bars. Two 
interesting points can be seen in the graphs on the left: ${}^{146}$Gd and ${}^{148}$Dy.
Both of these nuclei have neutron number of \(N= 82\), that is, one of the magic numbers. Here,
the theoretical predictions for the binding energies given by UNEDF1 and UNEDF2 are comparatively 
farther away from the experimental results. However, there is no visible increase in the standard 
deviation of the binding energy of these two nuclei. 
This suggests that the increased residual is due to underlying model deficiency, 
and not due to the parameter optimization procedure.

The calculated standard deviations of binding energies are found to be around 0.5--3.0\,MeV, 0.4--1.7\,MeV, 
and 0.3--1.5\,MeV for UNEDF0, UNEDF1 and UNEDF2, respectively. Even though the standard 
deviations have a magnitude of one thousandth of the total binding energy, 
the theoretical uncertainties are still far larger compared to  experimental precision,
which can be of the order of few keV's~\cite{(Kan12)}.
However, uncertainties of the UNEDF models have decreased after every model: The 
obtained standard deviation for UNEDF0 is larger compared to two later parameterizations.

The behavior of uncertainty is relatively smooth and the uncertainty of binding energy 
grows quickly when going towards neutron rich nuclei.
In addition, one can also see that uncertainties grow when going towards the other 
extreme, namely proton rich nuclei. 
This is an indication that isovector part of the EDFs is not as well constrained as the 
isoscalar part.

\begin{figure}[htb]
\centering
\includegraphics[width=0.8\textwidth]{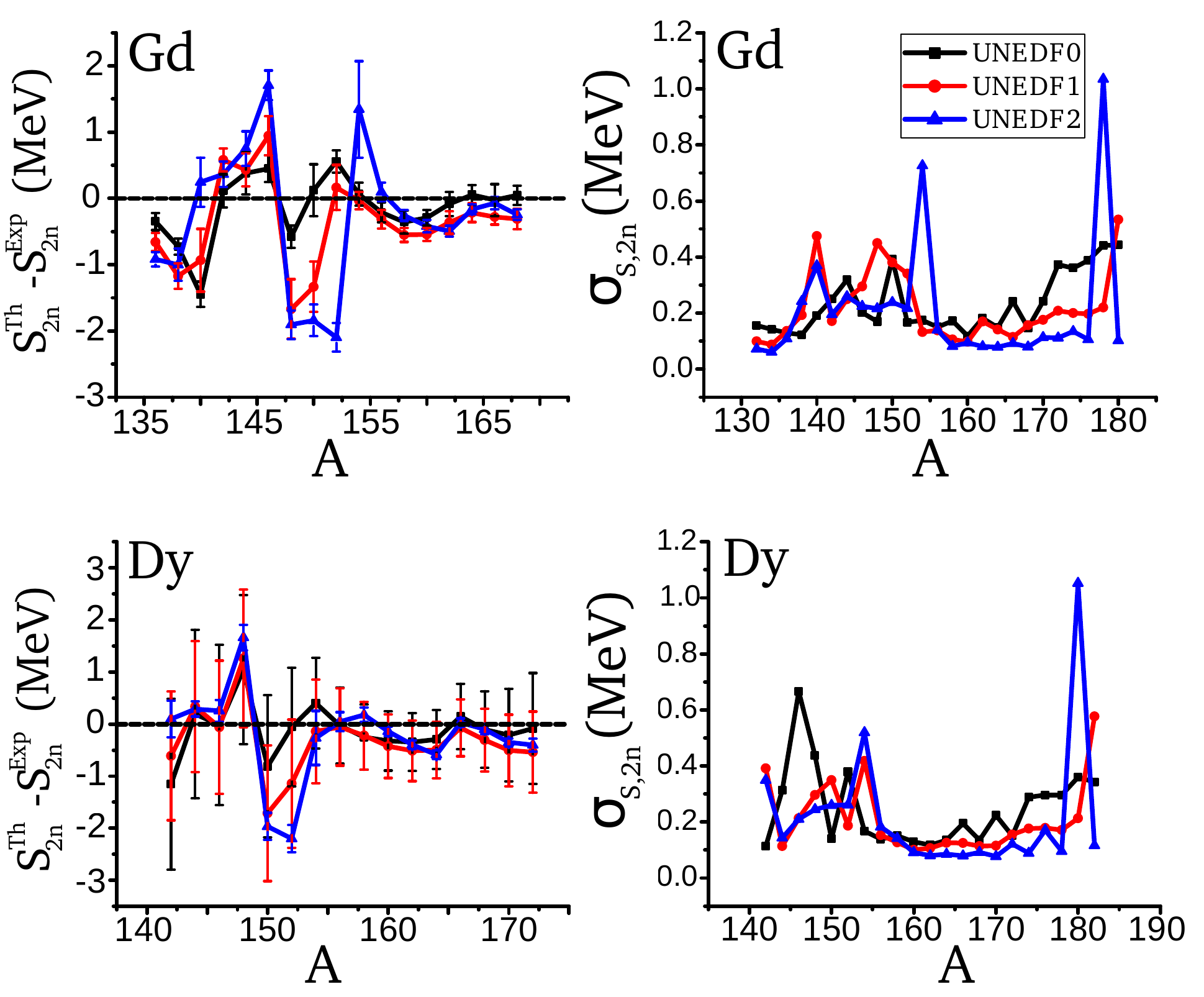}
\caption[The residuals for two-neutron separation energy of Gd ad Dy]{(Color online) 
Same as figure~\ref{fig:gddy_residuals} but for two-neutron separation energy \(S_{2n}\).}
\label{fig:gddy_sep_res}
\end{figure}

The residuals of the two-neutron separation energy, $S_{2\rm n}$, are shown in figure~\ref{fig:gddy_sep_res}. 
The two-neutron separation energies were calculated for even-even Dy and Gd isotopes.
Similarly to the previous figure, theoretical errors are marked as error bars in the graphs on the 
left hand side panels, and also given as a function of the mass number on the right hand side panels. 
The theoretical statistical error is calculated similarly, through finite differences of $S_{2\rm n}$
values to compute the derivatives. For neutron rich nuclei, all three 
parameterizations give essentially the same result for $S_{2\rm n}$, within the error bars. 
Otherwise, the latest UNEDF2 parameterization seems to differ most from the experimental 
results when compared to previous two parameterizations.

Since \(S_{2n}\) is defined as a difference of two binding energies, 
the partial derivatives are also calculated from energy difference between two nucleus.
As a consequence, some of the parameter uncertainties can cancel each other.
In particularly, the uncertainty coming from a relatively less constrained isovector part of 
the EDF is now partly canceled, resulting to more moderate uncertainty in the neutron rich region 
compared to the uncertainty of binding energy. Similar observation was also done at~\cite{(Gao13)}.

\begin{figure}[htb]
\centering
\includegraphics[width=0.99\textwidth]{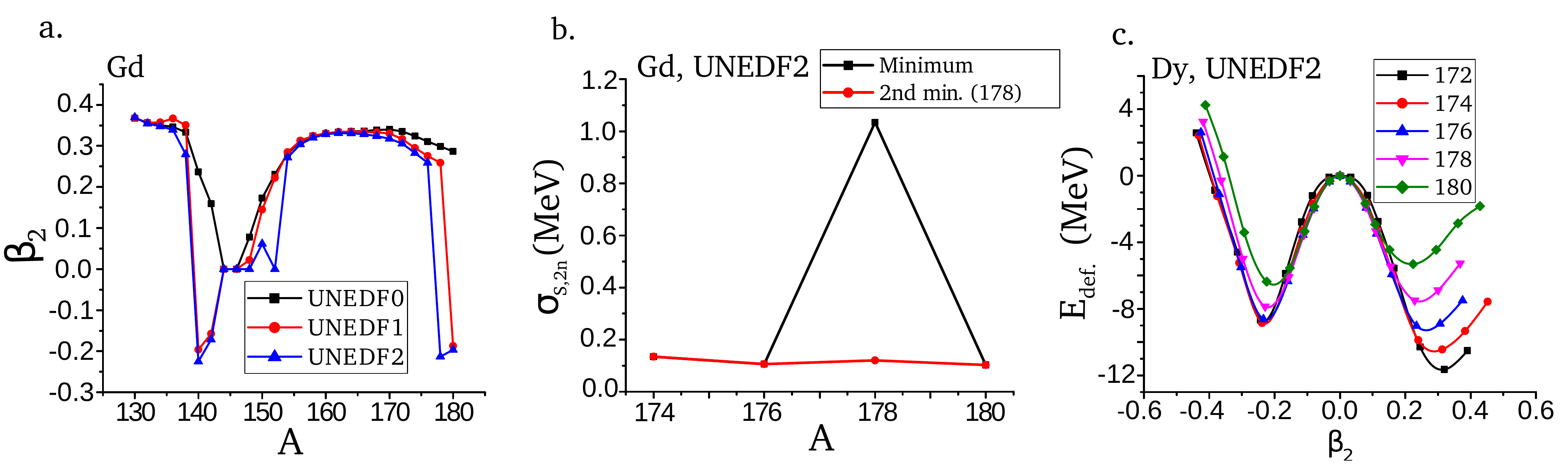}
\caption[Beta change]{(Color online) Panel a: Deformation parameter \(\beta_2\) as a function 
of mass number \(A\).
Panel b: Calculated statistical error of \(\sigma_{S,2n}\) when using either lowest or
secondary energy minimum of ${}^{178}$Gd.
Panel c: Deformation energies of five dysprosium isotopes as a 
function of deformation parameter \(\beta_2\).}
\label{fig:deform}
\end{figure}

One common feature for all UNEDF models is an existence of a few high peaks in the statistical 
error of \(S_{2n}\). These peaks are located mainly around the 
same neutron numbers for Gd and Dy isotopes. For instance, the two highest peaks given by UNEDF2 are 
located at the nuclei \({}^{178}\)Gd and \({}^{180}\)Dy. 
The explanation for all of the high peaks can be found in a sudden change in deformation. 
Figure~\ref{fig:deform}, in panel a, shows how deformation parameter \(\beta_2\) varies 
with mass number \(A\) for Gd. When comparing this to that of
\(\sigma_{S,2n}\), shown in figure~\ref{fig:gddy_sep_res}, one can notice similarity between
uncertainty peaks and large change in the deformation. If there is a significant 
difference in deformation between two consecutive even-even nuclei, this results to
a larger statistical error of two-neutron separation energy.

The relationship between \(\sigma_{S,2n}\) and a sudden large change of \(\beta_2\) can be tested by
looking at the secondary local minimum of the deformation energy landscape.
For the calculated Gd and Dy nuclei, there usually exists two energy minima, the oblate one 
and the prolate one, as shown in figure~\ref{fig:deform}, panel c.
By picking always the lowest minimum results to large statistical error of $S_{2n}$ when 
deformation has a large change between two even nucleus.
However, if one uses the secondary minimum, in which case the two nuclei appearing in the
expression of $S_{2n}$ have similar deformation compared to the each other, 
one obtains substantially smaller \(\sigma_{S,2n}\). 
In figure~\ref{fig:deform}, panel b, the black line describes the same 
peak at $A=178$ given by UNEDF2 as in figure~\ref{fig:gddy_sep_res}, 
calculated with the lowest energy minima.
The second (red) line corresponds to case where second minimum of \({}^{178}\)Gd was
used, resulting a much smaller \(\sigma_{S,2n}\) for \({}^{178}\)Gd.
Indeed, a large difference in deformation of involved nuclei seems to give large 
uncertainty on two-neutron separation energy. A possible explanation is considerably 
different shell structure between these two nuclei due to deformation. 
The largest impact on the extremely high peaks in \(\sigma_{S,2n}\) given by UNEDF2 is 
connected to the parameter \(C_{1}^{JJ}\), whereas for UNEDF1 the main contributors 
are \(M_s^*\), both \(C^{\rho \Delta J}\) together 
with \(C_0^{\rho \Delta\rho}\) and \(V_0^n\) parameters.

\begin{figure}[htb]
\centering
\includegraphics[width=0.8\textwidth]{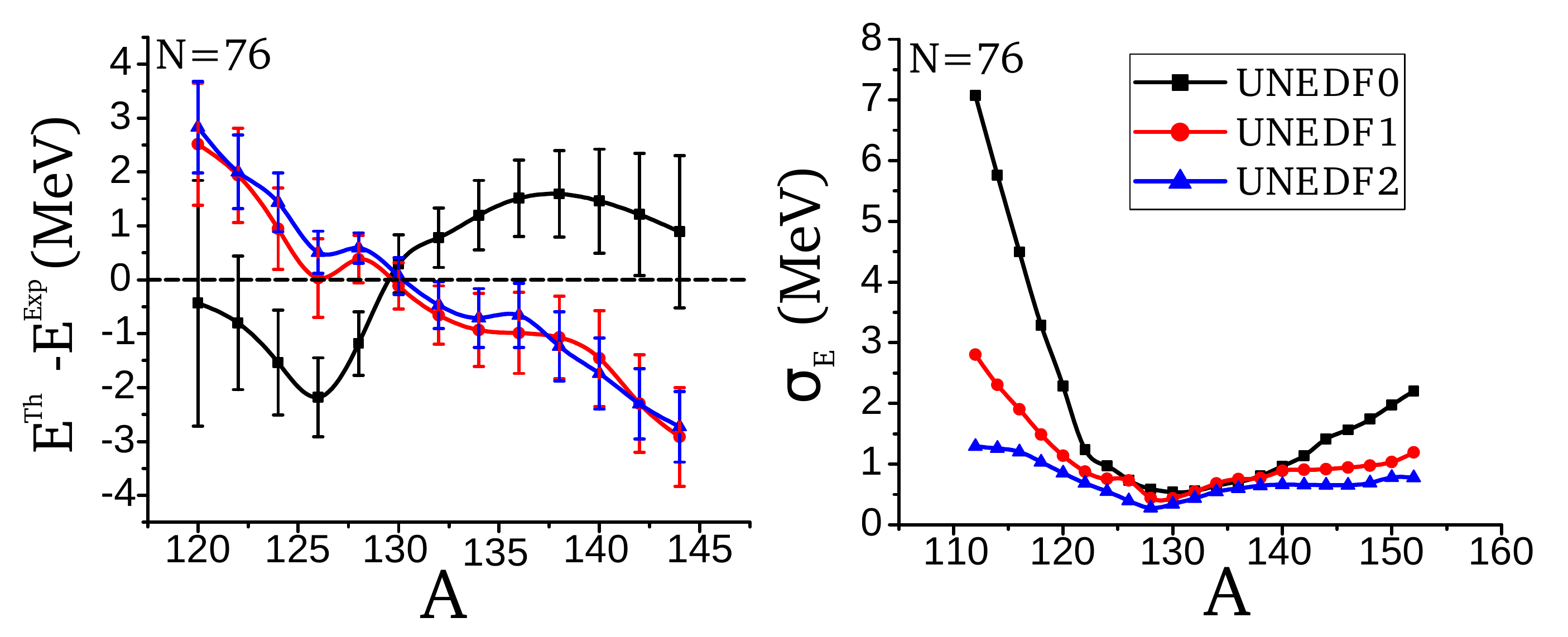}
\caption[Isotonic chain]{(Color online) 
Same as figure~\ref{fig:gddy_residuals}, but for the isotonic chain with the neutron number \(N=76\).}
\label{fig:isotonic_residuals}
\end{figure}

The binding energy residuals between theory and experiment for isotonic chain of \(N=76\) can be 
found in figure~\ref{fig:isotonic_residuals}. Only even-even nuclei are studied. The UNEDF1 and 
UNEDF2 parameterizations give rather similar results, but the binding energy behavior of UNEDF0 
parameterization is notably different. Compared to the UNEDF0 optimization procedure, in the 
optimization of UNEDF1 the same set of 12 EDF parameters were optimized but seven additional 
data points were included in the database and the center of mass correction was 
neglected~\cite{(Kor12)}. 
The other important remark is the fact that even though the trend of UNEDF1 and UNEDF2 models is 
incorrect -- the calculated binding energies for proton rich nuclei are getting further far away 
from the experimental ones when mass number increases -- the uncertainties do not nevertheless become larger.

\begin{figure}[htb]
\centering
\includegraphics[width=0.8\textwidth]{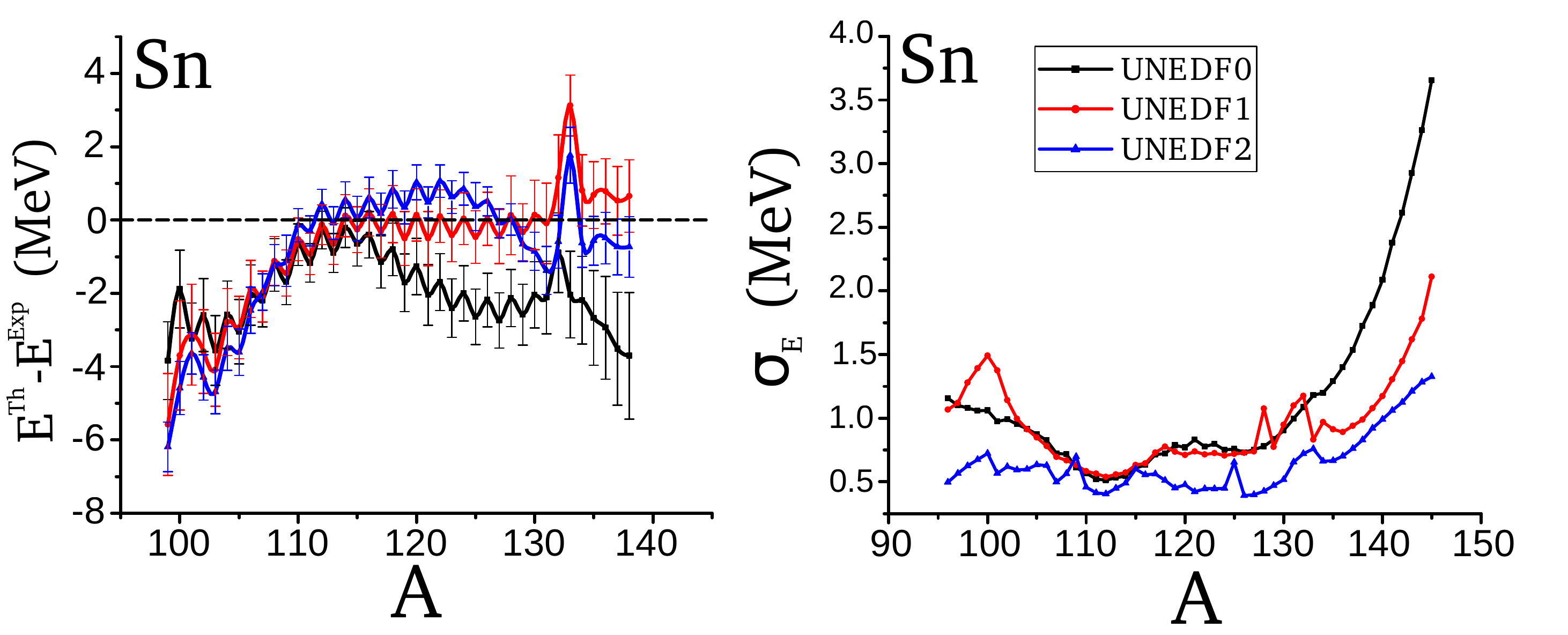}
\caption[Binding energy residual for Sn]{(Color online) 
Same as figure~\ref{fig:gddy_residuals} but for even-even and even-odd Sn isotopes.}
\label{fig:sn_residual}
\end{figure}

We have also considered the statistical error of odd-even nuclei.
The binding energy residual for Sn isotopic chain was computed by using all three UNEDF models.
The results are shown in figure~\ref{fig:sn_residual}. 
The odd-even nuclei were calculated by using the quasi\-particle blocking
procedure with the equal filling approximation~\cite{(Sch10)}.
The same blocking configuration, which corresponded the lowest energy with unshifted parameterization,
was used for calculation of all partial derivatives.
The same set of results for even Sn isotopes, with UNEDF0, was calculated in~\cite{(Gao13)}.
The results show that the binding energy residuals of even-even nuclei are relatively greater
compared to those of odd nuclei.
On this account, the binding energy residuals stagger between the odd and even nuclei.
Nevertheless, there are no visible odd-even effects in the standard deviations of binding energy.
This can be explained by the lack of time-odd part in the used EDFs.

\subsection{Uncertainty of \(Q_{2,p}\) and proton matter radius}

\begin{figure}[htb]
\centering
\includegraphics[width=0.99\textwidth]{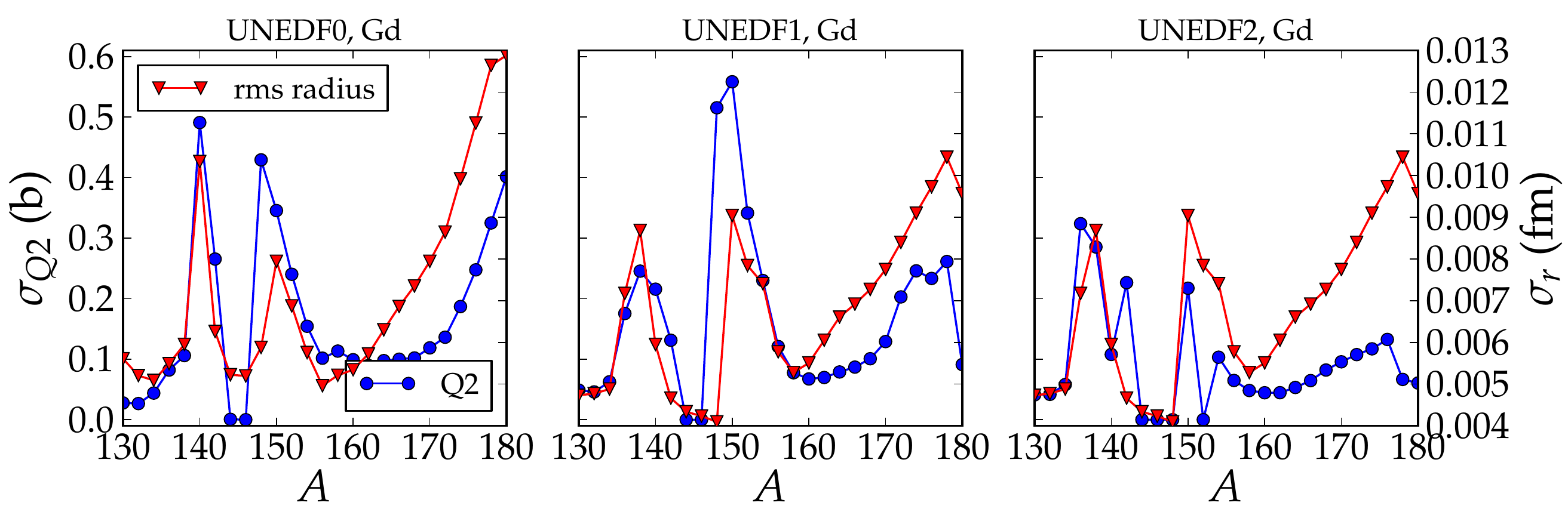}
\caption[]{(Color online) The standard deviation of proton quadrupole moment \(Q_{2,p}\) 
and proton rms radius \(r_{p,rms}\) for Gd isotopic chain with all three UNEDF models.
The scale of \(\sigma_{Q2}\) can be read on the left side 
and the scale of \(\sigma_{r}\) on the right side of the figure } 
\label{fig:AQ2}
\end{figure}

The standard deviation of proton matter quadrupole moment \(Q_{2,p}\) and proton matter 
root-mean-square (rms) radius \(r_{p,rms}\) for
all three UNEDF models is shown in figure~\ref{fig:AQ2}. The scale of \(\sigma_{Q2}\) can be
read on the left side and the scale of \(\sigma_{r}\) on the right side of the figure.
As expected, the uncertainty of these two observables behaves similarly and is strongly correlated.
High values of uncertainty are located in deformed nuclei next to spherical nuclei, due to
soft deformation energy landscape with respect of quadrupole deformation.

Despite the general strong correlation between \(\sigma_{r}\) and \(\sigma_{Q2}\), with UNEDF1 and UNEDF2, there are 
a few points of \(Q_{2,p}\) which differ from the major trend. The vanishing
uncertainty of \(Q_{2,p}\) in \({}^{152}\)Gd is explained by the fact 
that \({}^{152}\)Gd is predicted to be spherical by UNEDF2 EDF. This can be seen
in figure~\ref{fig:deform}, panel a.
The divergent uncertainty of \(Q_{2,p}\) in \({}^{180}\)Gd (\({}^{178}\)Gd, 
\({}^{180}\)Gd) in UNEDF1 (UNEDF2) is related to
the change of sign in quadrupole moment and deformation parameter \(\beta_2\)
which is shown also in figure~\ref{fig:deform}, panel a.
Most of the nuclei are predicted to be prolate
by UNEDF1 and UNEDF2, but above-mentioned nuclei are predicted to be oblate, resulting to
rapid changes of the statistical uncertainty of \(Q_{2,p}\) for Gd isotopic chain. However,
\({}^{140}\)Gd and \({}^{142}\)Gd isotope are also predicted to be oblate by UNEDF1 and UNEDF2 EDFs, 
but there is no significant effect in the uncertainties given by UNEDF1. There is no
oblate-shaped nuclei among the Gd isotopes calculated with UNEDF0 EDF.

In addition to the large uncertainties next to the spherical nuclei, there is also
another visible trend in the uncertainties. 
Similarly like with the uncertainties of the binding energy, when going towards the neutron 
rich nuclei, the uncertainties of \(Q_{2,p}\) and
\(r_{p,rms}\) increases systematically. The same behavior is also followed 
with the uncertainties of the neutron matter radius.

\subsection{Contributions of the model parameters}

\begin{figure}[htb]
\centering
\includegraphics[width=0.95\textwidth]{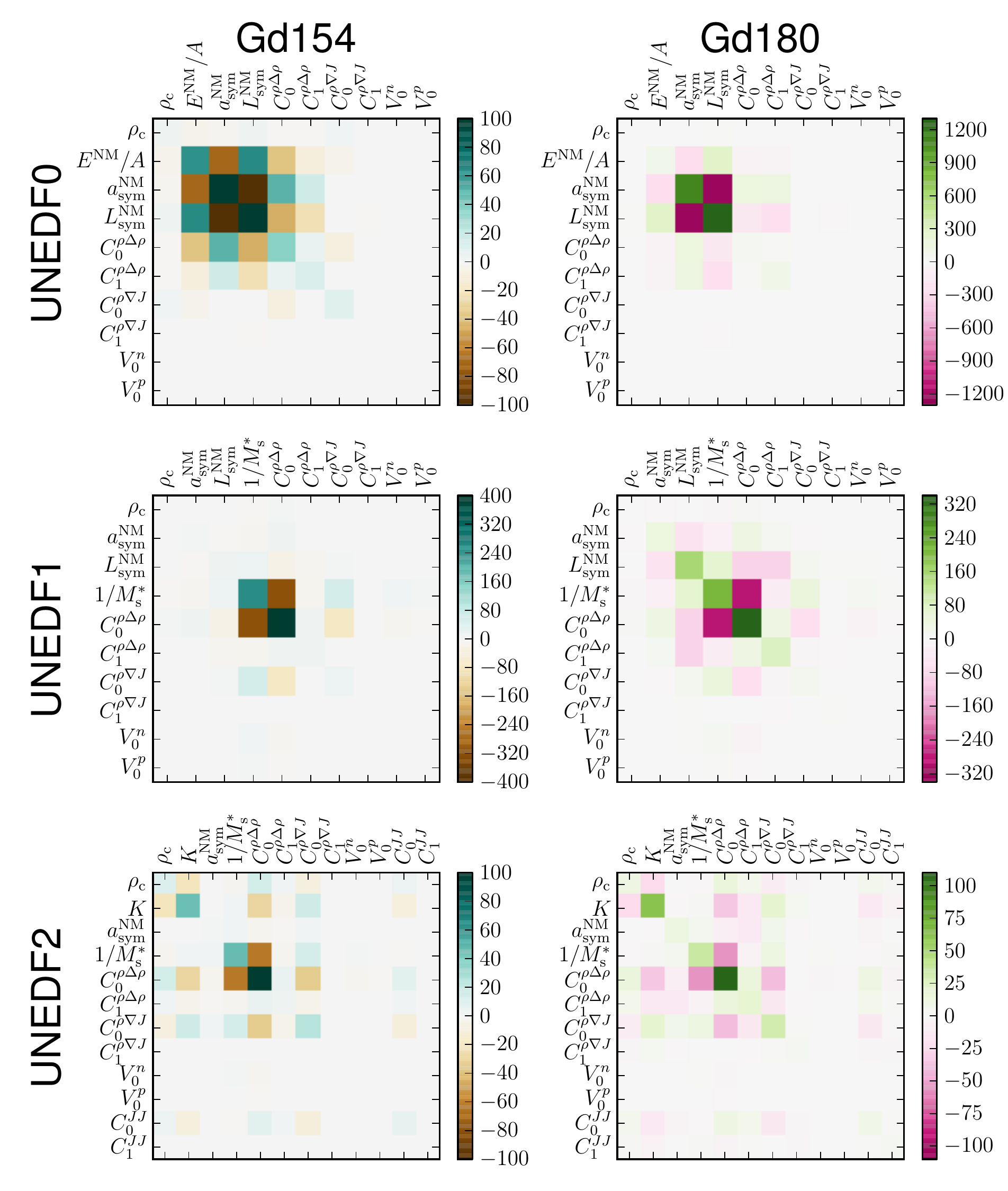}
\caption[]{(Color online) Individual contributions of the model parameters (\(x_i, x_j\)), in 
the total sum of equation (\ref{eq:sigma2}), for the uncertainty of binding energy 
in $^{154}$Gd and $^{180}$Gd isotopes and for all three UNEDF models.
The contributions are in units of MeV$^2$.} 
\label{fig:colormatr}
\end{figure}

One of the goals of present work is to study contributions of the model parameters to the 
total error budget of binding energy. The most elementary way to represent the contributions of 
the model parameters to the total uncertainty is by listing component matrix. Here, every single small color 
square in the matrix represents the value of one particular cross contribution coming from 
parameters (\(x_i, x_j\)) to the total sum of equation~(\ref{eq:sigma2}). 
The component matrices for the binding energy uncertainties of \({}^{154}\)Gd 
and \({}^{180}\)Gd  are shown in figure~\ref{fig:colormatr}. 
Some of these components have negative sign, due to a negative partial derivative or a negative covariance matrix element. 
The total squared standard deviation is always, nevertheless, a positive number.
One should also bear in mind that the contribution of 
a parameter to the standard deviation is visible only if this parameter was included in the 
sensitivity analysis, as mentioned before. 

For \({}^{154}\)Gd, which has one of the smallest statistical error of binding energy with all UNEDF
parameterizations, the total error budget with UNEDF0 and UNEDF2 EDFs consist of several 
components. The total error budget with UNEDF1 is simpler, and mainly coming from 
\(1/M_s^*\) and \(C_0^{\rho \Delta \rho}\) parameters. On the other hand, the 
uncertainty budget for the 
neutron rich nucleus \({}^{180}\)Gd splits into several pieces when going from the 
oldest parameterization to the latest one. The uncertainty of UNEDF0 is affected by 
a couple of parameters, mainly by \(a^{\rm NM}_{\rm sym}\) and \(L^{\rm NM}_{\rm sym}\).
The most contributing parameters to the uncertainty of
UNEDF1 are \(1/M_s^*\) and \(C_0^{\rho \Delta \rho}\), whereas in the case of UNEDF2
the \(C_0^{\rho \Delta \rho}\) parameter have slightly greater contribution. Even 
though one parameter has slightly greater contribution, the uncertainty of binding energy 
for neutron rich nuclei is relatively widely spread among the model parameters of UNEDF2.

The component matrix representation shows explicitly how different model parameters contribute 
to the total error budget pairwise. Unfortunately, this representation requires a lot of space.
By considering a summed contribution of one row (or, equivalently, one column) of the component matrix,
we can represent the error budget as a stacked histograms for each isotope.
We refer this once summed contribution as a \emph{row contribution} of a parameter \(x_i\). Here,
components of the total error are calculated by summing over one index in equation~(\ref{eq:sigma2}),
resulting the total squared standard deviation being then a sum of all row contributions.

\begin{figure}[htb]
\centering
\includegraphics[width=0.75\textwidth]{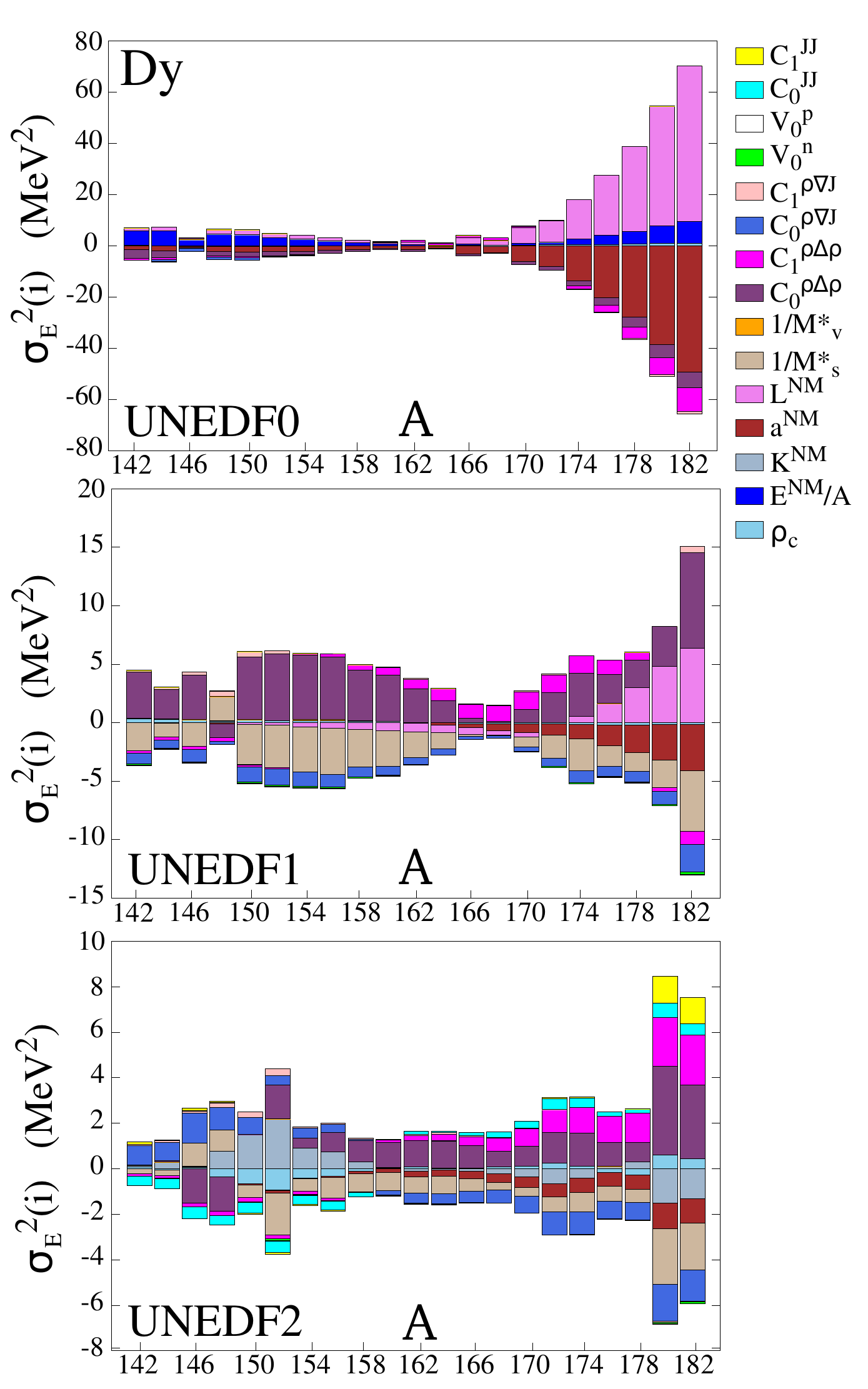}
\caption[]{(Color online) Error budget of \(\sigma_E^2\) for even-even Dy isotopes with
UNEDF0, UNEDF1 and UNEDF2 EDFs in the row contribution representation. The 
contribution of a summed up row is indicated with corresponding model parameter name.} 
\label{fig:dycomponents}
\end{figure}

\begin{figure}[htb]
\centering
\includegraphics[width=0.75\textwidth]{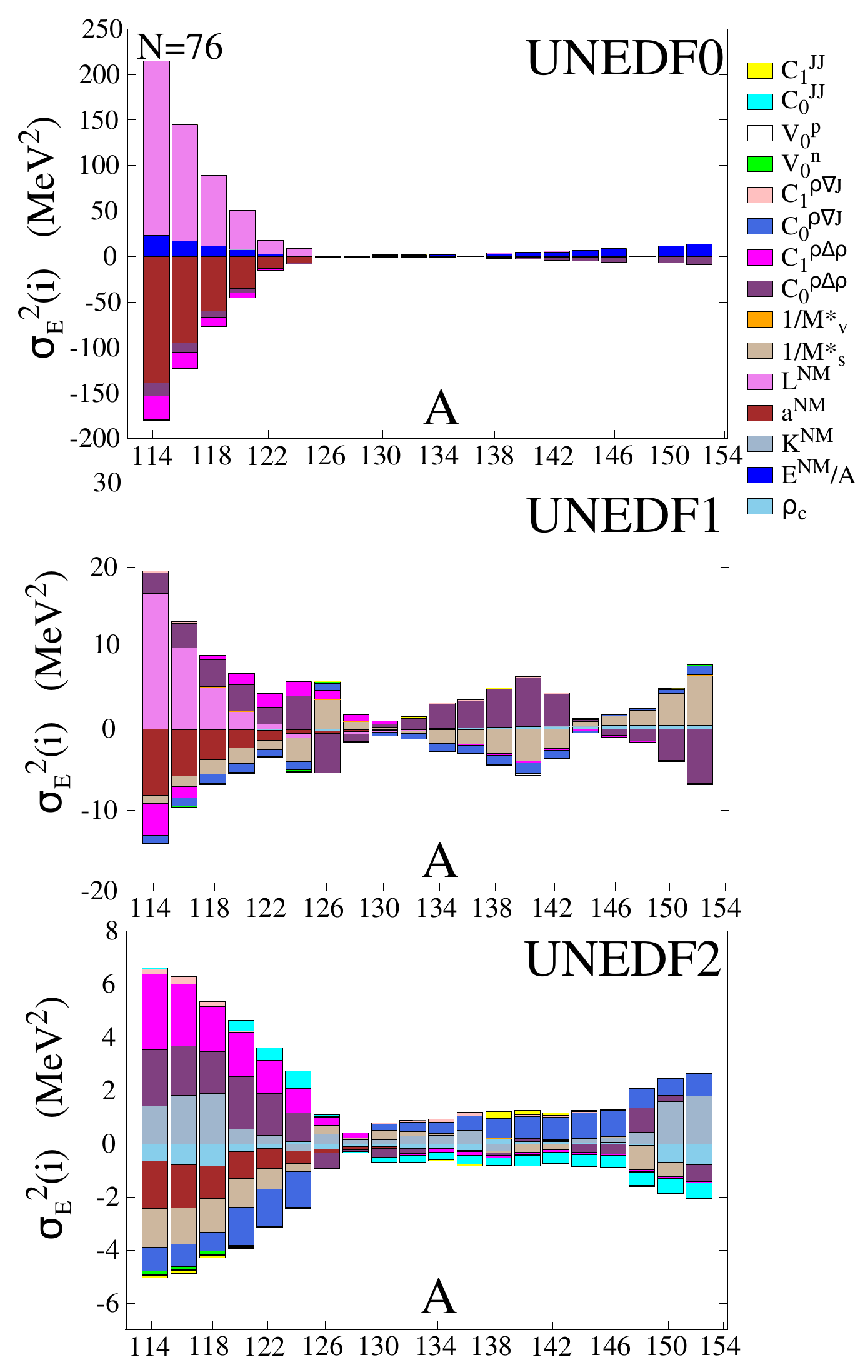}
\caption[]{(Color online) Same as figure~\ref{fig:dycomponents} but for the isotonic chain of \(N=76\).}
\label{fig:isotonic_components}
\end{figure}

The results for the row contributions are shown in figure~\ref{fig:dycomponents} 
for the binding energies of even Dy isotopes and in figure~\ref{fig:isotonic_components}
for the binding energies of even \(N=76\) isotones.
The error budget for UNEDF0 binding energy is mainly composed of only a few contributing rows. 
For nuclei close to the valley of stability, two dominant sources of uncertainty 
are the rows \(E^{\rm NM}/A\) and \(C_0^{\rho \Delta \rho}\) parameters, whereas in the neutron rich 
Dy isotopes the rows of \(L_{\rm sym}^{\rm NM}\) and \(a_{\rm sym}^{\rm NM}\) dominate. 
In other words, model parameters related to symmetry energy become more important 
with neutron rich nuclei. It was found earlier that \(L_{\rm sym}^{\rm NM}\) has also a strong 
impact on the statistical error of neutron root-mean-square radii and neutron skin 
thickness~\cite{(Gao13),(Kor13)}.
The most dominant sources of uncertainty are the same for isotopic and isotonic chains. 

Contrary to UNEDF0, the error budget of latter two parameterizations is more split
among the various different row contributions in neutron rich nuclei. 
Generally speaking, the rows connected to \(a^{\rm NM}_{\rm sym}\), both \(C_t^{\rho \Delta \rho}\) parameters, 
\(C_0^{\rho \nabla J}\), and \(1/M^*_s\) have a significant impact on the total 
error budget with UNEDF2. 
It should be noticed that the correlation between \(C_0^{\rho \Delta \rho}\) and \(1/M^*_s\) 
is strong: In principle, if one can reduce the uncertainty on \(C_0^{\rho \Delta \rho}\), 
it should also reduce the uncertainty of \(1/M^*_s\). 
The isovector parameters, for their part, are more difficult to constrain, but they impact 
on the stability of the functional: For instance \(C_1^{\rho \Delta \rho}\) is the 
coupling constant of the gradient term, which has been found to trigger scalar-isovector 
instabilities~\cite{(Hel13)}. 

%--------------------------
% Eigenvector representation

\begin{figure}[htb]
\centering
\includegraphics[width=0.99\textwidth]{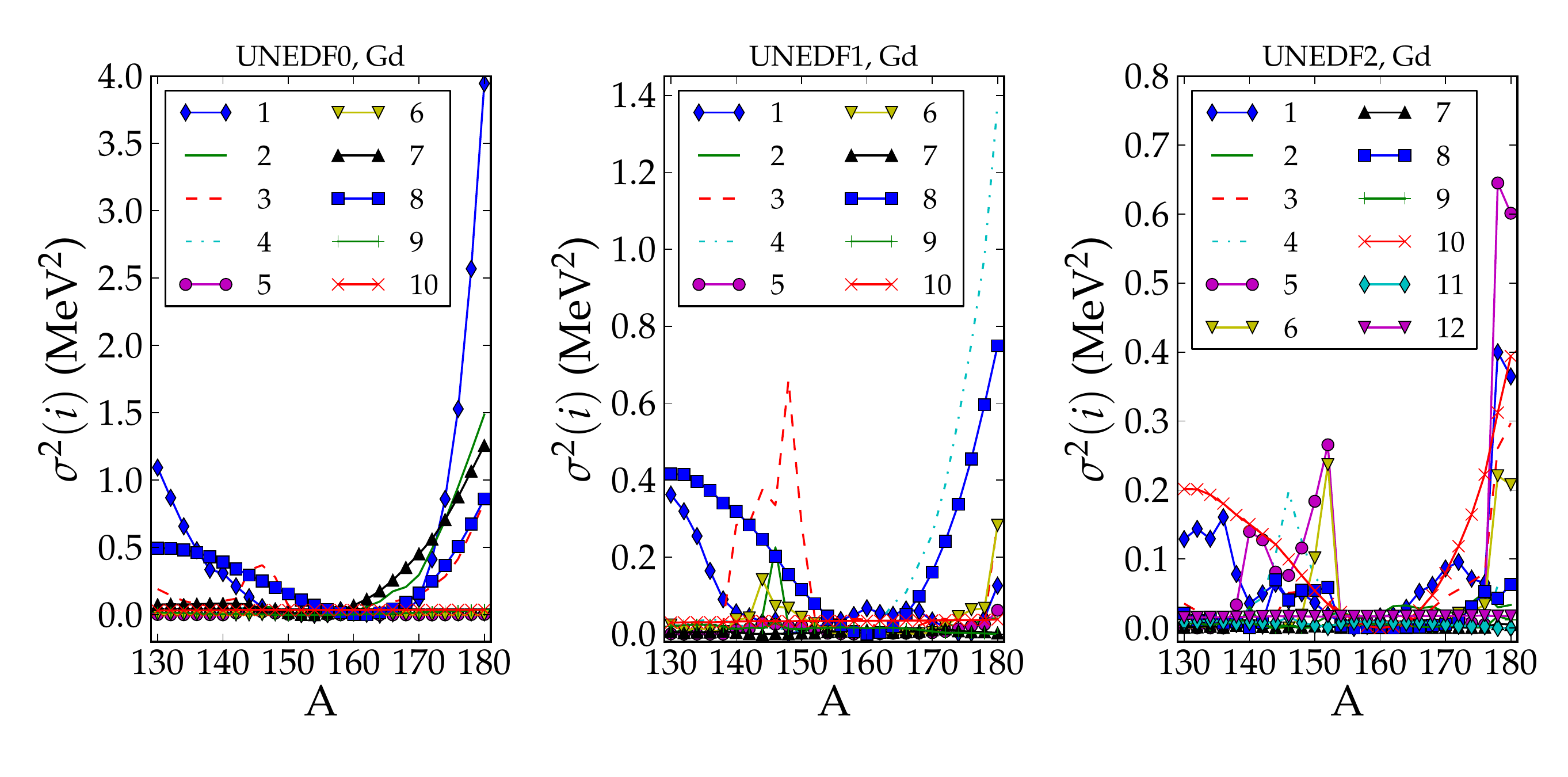}
\caption[]{(Color online) The uncertainties of binding energies in the eigenmode formalism. Each curve represents
the contribution of one eigenvector. The eigenvectors  are different 
for each UNEDF parameterization, even though symbols and numbering is the same.} 
\label{fig:eigens}
\end{figure}

Lastly, we represent the uncertainties of binding energies in the eigenmode 
formalism in figure~\ref{fig:eigens}.
The eigenvectors are listed in descending order of eigenvalues.
The eigenvectors and eigenvalues are not shown here - it is not a laborious task to
diagonalize covariance matrices given in references~\cite{(Kor10), (Kor12), (Kor14)}.
Basically, a small eigenvalue means that the linear combination of model parameters described by this 
eigenvector is well constrained. If the eigenvalue is large, the corresponding
eigenvector is poorly constrained. The eigenvector representation does not directly tell
about the model parameters themselves, but describes how the uncertainty 
propagates from a certain linear combinations of the model parameters, instead.
As we can see, e.g. for neutron rich
nuclei, only five eigenvectors of UNEDF0 contribute significantly to 
the total error budget. The eigenvector having the greatest eigenvalue, and thus being least constrained,
has also the biggest contribution to the error budget among the neutron and proton rich
nuclei.
In the case of UNEDF1, mainly two eigenvectors 
contribute to the total error budget of a given nucleus, whereas there are 5 significant
contributors with UNEDF2 EDF. Interestingly, with UNEDF1, one can see many contributing
eigenvectors at deformation transition region around $A=140$ -- $152$.

We can also investigate how different model parameters contribute to the uncertainty, in terms of
components of one particular eigenvector. For example, with UNEDF0, the first eigenvector has 
the biggest contribution with neutron rich nuclei. When looking at the individual components of this
eigenvector, the main contributing model parameters are \(a^{\rm NM}_{\rm sym}\) and \(L^{\rm NM}_{\rm sym}\).
With the second largest contributing eigenvector, the contribution of \(a^{\rm NM}_{\rm sym}\) 
 is the largest one and with third most contributing eigenvector,
the parameters
\(C_0^{\rho \Delta \rho}\), \(a^{\rm NM}_{\rm sym}\) and \(E^{\rm NM}/A\)
 are the most important ones.
With  UNEDF1, the uncertainty of binding energy for neutron rich
nuclei comes from \(a^{\rm NM}_{\rm sym}\), \(L^{\rm NM}_{\rm sym}\) and \(C_0^{\rho \Delta \rho}\) 
model parameters when considering the most contributing eigenvector, and from 
\(a^{\rm NM}_{\rm sym}\) and \(\rho_c\) when considering eigenvector having the second largest contribution. 
On the other hand, with UNEDF2, the contributions are more split among various model parameters. 
Largest contributing components of two most important
eigenvectors consist of several parameters (more than five). The uncertainty coming from
the third most contributing eigenvector consists almost entirely of \(a^{\rm NM}_{\rm sym}\)
and \(\rho_c\) model parameters.

In the end of this section, we can conclude that three different methods used in this 
study, namely the component matrix representation, the histogram representation of row contributions,
and the eigenmode method, are in support of each other, with each one having their own advantage.

\section{Conclusions}
\label{sec:concl}

In the present work we have calculated statistical errors of the UNEDF models 
for binding energy, two-neutron separation energy, proton quadrupole 
moment and proton rms radius by using information about
the covariance matrix of the model parameters.
The standard deviation has been interpreted as a statistical error. 
We have also quantified the contributions of each model parameters to the total error budget
by using three different methods
and checked if there are any visible odd-even effects in the uncertainties of the UNEDF models.
We presented our results for the error budget by using the component matrix representation, the row contribution
representation, and finally, by using the eigenmode method.
We found out that the standard deviation of binding energy grows 
quickly when going away from the valley of stability
towards proton rich or neutron rich nuclei. Similarly, uncertainties of proton quadrupole moment
\(Q_{2,p}\) and proton rms radius \(r_{rms,p}\) increase rapidly among the neutron rich
nuclei as a function of mass number. 
That is to say, the predictive power of UNEDF models becomes weaker when extrapolating
further away from known nuclei to experimentally unknown region.
For the Sn isotopic chain, even though there exists odd-even staggering in the residuals of binding energies, 
no visible odd-even effect was seen in the related errors.

The error budget of the UNEDF parameterizations becomes more evenly split among various
model parameters with UNEDF1 and UNEDF2 parameterizations in neutron rich nuclei. 
This can be seen by using any of the three methods mentioned above.
The most dominant contributors to the error budget of neutron rich nuclei with UNEDF0 
were \(L^{\rm NM}_{\rm sym}\) and \(a^{\rm NM}_{\rm sym}\) parameters, that is to say, coefficients related to the symmetry energy. 
In the case of UNEDF1, \(L^{\rm NM}_{\rm sym}\) and \(a^{\rm NM}_{\rm sym}\) still have a significant contribution
and, in addition, the role of \(C_0^{\rho \Delta \rho}\) and \(1/M^*_s\) becomes important.
With UNEDF2, the role of other model parameters becomes even more prominent.

A comparison of the binding energy standard deviations to the residuals seems to point out that the underlying 
theoretical model is missing some important physics.
One clear indication of this is the increase of the binding energy residuals close to the semi-magic 
nuclei. However, calculated standard deviation does not usually reflect this kind
of behavior. Similar observation was also done in~\cite{(Gao13)}.
Also, the systematically incorrect trend of UNEDF1 and UNEDF2 with the binding energy residual in the $N=76$
isotonic chain does not appear in theoretical uncertainties, as shown in
figure~\ref{fig:isotonic_residuals}.

The trend in the UNEDF statistical errors is such that the calculated standard deviation decreases
as more data is included in the optimization procedure.
From a statistical point of view, the larger uncertainty of the isovector parameters reflects
to the increasing uncertainty when going towards the both isospin extremes. 
Unfortunately, new data may not always help to reduce uncertainties.
A sophisticated parameter optimization within Bayesian framework showed that new data points
on nuclear binding energies at neutron rich region were not able provide a better constraint 
for the UNEDF1 model parameters~\cite{(McD15)}.
On the other hand, data on the neutron skin thickness could potentially help to reduce 
uncertainties related on isovector model parameters~\cite{(Rei10)}.

As was concluded for the UNEDF2 parameterization, the limits of standard Skyrme-EDFs have been 
reached and novel approaches are called for~\cite{(Kor14)}. 
Information about the shortcomings and uncertainties of present models will provide valuable input for
development of novel EDF models. A comprehensive sensitivity analysis of a novel EDF parameterization
is essential when addressing its predictive power.

\section*{Acknowledgements}
T.H. thanks Karim Bennaceur and Nicolas Schunck for great advice and remarks. This work was supported by the 
Academy of Finland under the Centre of Excellence Programme 2012-2017 (Nuclear and 
Accelerator Based Physics Programme at JYFL) and under the FIDIPRO programme. 
T.H. was also supported by a grant (55151927) of the Finnish Cultural Foundation, 
North Karelia Regional Fund.

\clearpage

%\newpage
\appendix
\section{Appendix: Correlation matrices} \label{App:AppendixA}
%\subsection{UNEDF0}
\setcounter{section}{1}

\begin{table}[!h]
\caption[]{Correlation matrix and standard deviations of the UNEDF0 parameter set~\cite{(Kor10)}. 
The values are rounded. The units are following: \(\rho_{\rm c} \) is in \($fm$^{-3}\); \(E^{\rm NM}/A \), 
\(a_{\rm sym}^{\rm NM}\) and \(L_{\rm sym}^{\rm NM}\) are in MeV; \(C_{t}^{\rho \Delta \rho}\) 
and \(C_{0}^{\rho \nabla J}\) are in MeV \($fm$^{-5}\), and \(V_0^{n,p}\) are in MeV \($fm$^{-3}\). }
\centering
\scalebox{0.9}{
\begin{tabular}{ccccccccccc}
\br
 & \(\rho_{\rm c} \) & \(E^{\rm NM}/A \) &  \(a_{\rm sym}^{\rm NM}\) & \(L_{\rm sym}^{\rm NM}\) &  \(C_{0}^{\rho \Delta \rho} \) & \(C_{1}^{\rho \Delta \rho} \) & \(V_0^{n}\) & \(V_0^{p}\)   & \(C_{0}^{\rho \nabla J}\) &  \(C_{1}^{\rho \nabla J}\)  \\
\mr
\(\rho_{\rm c} \) &     1.00 &  &&&&&&&& \\
\(E^{\rm NM}/A \) &  -0.28 &  1.00 &&&&&&&& \\
\(a_{\rm sym}^{\rm NM}\) &  -0.10 & -0.88 &  1.00 &&&&&&& \\
\(L_{\rm sym}^{\rm NM}\) &  -0.17 & -0.80 &  0.97 &  1.00 &&&&&& \\
\(C_{0}^{\rho \Delta \rho} \) &   0.09 &  0.80 & -0.81 & -0.74 &  1.00 &&&&& \\
\(C_{1}^{\rho \Delta \rho} \) &   0.20 &  0.35 & -0.47 & -0.66 &  0.23 &  1.00 &&&& \\
\(V_0^{n}\) &   0.02 &  0.21 & -0.23 & -0.25 &  0.23 &  0.23 &  1.00 &&& \\
\(V_0^{p}\) &  -0.13 & -0.42 &  0.52 & 0.56 & -0.29 & -0.45 & -0.14 & 1.00 &&\\
\(C_{0}^{\rho \nabla J}\) &   0.37 & -0.14 &  0.02 & -0.00 &  0.44 & -0.02 &  0.09 & 0.16 & 1.00 & \\
\(C_{1}^{\rho \nabla J}\) &  -0.06 & -0.18 &  0.27 & 0.33 & -0.38 & -0.20 & -0.01 & 0.00 & -0.37 &  1.00  \\
\mr
\(\sigma\) & 0.001 & 0.055 & 3.058 & 40.037 & 1.697 & 56.965 & 2.105 & 3.351 & 3.423 & 29.460 \\
\br
\end{tabular}
}
\label{tab:unedf0corr}
\end{table}

%\subsection{UNEDF1}
\begin{table}[!h]
\caption[]{Same as table~\ref{tab:unedf0corr} but for UNEDF1. \(L_{\rm sym}^{\rm NM}\) is in 
units of MeV and \(1/M^*_{\rm s}\) is unitless.}
\centering
\scalebox{0.9}{
\begin{tabular}{ccccccccccc}
\br
 & \(\rho_{\rm c} \) & \(a_{\rm sym}^{\rm NM}\) &  \(L_{\rm sym}^{\rm NM}\) & \(1/M^*_{\rm s}\) &  \(C_{0}^{\rho \Delta \rho} \) & \(C_{1}^{\rho \Delta \rho} \) & \(V_0^{n}\) & \(V_0^{p}\) & \(C_{0}^{\rho \nabla J}\) &  \(C_{1}^{\rho \nabla J}\)    \\
\mr
\(\rho_{\rm c} \) &     1.00 &  &&&&&&&& \\
\(a_{\rm sym}^{\rm NM}\) &  -0.35 &  1.00 &&&&&&&& \\
\(L^{\rm NM} \) & -0.14 & 0.71 &  1.00 &&&&&&& \\
\(1/M^*_{\rm s}\) &  0.32 & 0.23 &  0.36 &  1.00 &&&&&& \\
\(C_{0}^{\rho \Delta \rho} \) &   -0.25 &  -0.25 & -0.35 & -0.99 &  1.00 &&&&& \\
\(C_{1}^{\rho \Delta \rho} \) &   -0.06 &  -0.15 & -0.77 & -0.22 &  0.19 &  1.00 &&&& \\
\(V_0^{n}\) &   -0.32 & -0.22 &  -0.36 & -0.99 &  0.98 & 0.22 &  1.00 &&& \\
\(V_0^{p}\) &  -0.33 & -0.18 &  -0.29 & -0.97 & 0.97 & 0.15 & 0.96 &  1.00 && \\
\(C_{0}^{\rho \nabla J}\) &   -0.14 &  -0.20 & -0.32 & -0.86 &  0.91 &  0.22 &  0.85 & 0.84 & 1.00 & \\
\(C_{1}^{\rho \nabla J}\) &  0.05 & -0.17 &  -0.13 & -0.10 & 0.07 & 0.21 &  0.10 &  0.07 & -0.03 & 1.00 \\
\mr
\(\sigma\) & 0.0004 & 0.604 & 13.136 & 0.123 & 5.361 & 52.169 & 18.561 & 13.049 & 5.048 & 23.147 \\
\br
\end{tabular}
}
\label{tab:unedf1corr}
\end{table}

%\newpage
%\subsection{UNEDF2}
\begin{table}[!h]
\caption[]{Same as table~\ref{tab:unedf0corr} but for UNEDF2. The parameters 
\(C_{1}^{\rho \nabla J}\) are in units of MeV \($fm$^{-3}\).}
\centering
\scalebox{0.80}{
\begin{tabular}{ccccccccccccc}
\br
 & \(\rho_{\rm c} \) & \(K_{\rm sym}^{\rm NM}\) &  \(a_{\rm sym}^{\rm NM}\) & \(1/M^*_{\rm s}\) &  \(C_{0}^{\rho \Delta \rho} \) & \(C_{1}^{\rho \Delta \rho} \) & \(V_0^{n}\) & \(V_0^{p}\) & \(C_{0}^{\rho \nabla J}\) &  \(C_{1}^{\rho \nabla J}\) & \(C_{0}^{JJ}\) & \(C_{1}^{JJ}\)   \\
\mr
\(\rho_{\rm c} \) &     1.00 &  &&&&&&&&&& \\
\(K_{\rm sym}^{\rm NM}\) &  -0.97 &  1.00 &&&&&&&&&& \\
\(a_{\rm sym}^{\rm NM}\) & -0.07 & -0.03 &  1.00 &&&&&&&&& \\ 
\(1/M^*_{\rm s}\) &  0.08 & -0.05 &  -0.24 &  1.00 &&&&&&&& \\
\(C_{0}^{\rho \Delta \rho} \) &   -0.43 &  0.43 & 0.22 & -0.89 &  1.00 &&&&&&& \\
\(C_{1}^{\rho \Delta \rho} \) &   -0.42 &  0.37 & 0.83 & -0.17 &  0.31 &  1.00 &&&&&& \\
\(V_0^{n}\) &   -0.06 & 0.02 &  0.27 & -0.96 &  0.85 & 0.17 &  1.00 &&&&& \\
\(V_0^{p}\) &  -0.09 & 0.05 &  0.21 & -0.89 & 0.80 & 0.14 & 0.86 &  1.00 &&&& \\
\(C_{0}^{\rho \nabla J}\) &   -0.51 &  0.50 & 0.34 & -0.40 &  0.68 &  0.55 &  0.36 & 0.34 & 1.00 &&& \\
\(C_{1}^{\rho \nabla J}\) &  -0.31 & 0.29 &  -0.19 & -0.00 & 0.04 & 0.18 &  -0.07 &  -0.02 & 0.14 & 1.00 && \\
\(C_{0}^{JJ}\) &   0.56 &  -0.55 & -0.26 & 0.05 &  -0.35 &  -0.53 &  -0.02 & -0.02 & -0.88 & -0.35 & 1.00 &  \\
\(C_{1}^{JJ}\) &  0.36 & -0.35 &  0.13 & -0.23 & 0.16 & -0.14 &  0.29 &  0.25 & -0.02 & -0.57 & 0.29 & 1.00 \\
\mr
\(\sigma\) & 0.001 & 10.119 & 0.321 & 0.052 & 2.689 & 24.322 & 8.353 &  6.792 & 5.841 & 15.479 & 16.481 & 17.798\\
\br
\end{tabular}
}
\label{tab:unedf2corr}
\end{table}

%\newpage
\clearpage
\section{Appendix: Values of the parameters and used finite differences} \label{App:AppendixB}

\begin{table}[!h]
\caption[Finite differences]{Rounded values \(x_i\) for each UNEDF parameterization and 
the corresponding used finite differences \(\Delta x_i\) used in the derivatives. 
The units are the same as in tables~\ref{tab:unedf0corr},~\ref{tab:unedf1corr} and~\ref{tab:unedf2corr}.}
\centering
\scalebox{0.9}{
\begin{tabular}{lrlrlrl}
\br
& \multicolumn{2}{c}{UNEDF0} & \multicolumn{2}{c}{UNEDF1} & \multicolumn{2}{c}{UNEDF2} \\
%\cmidrule(r){2-3} 
%\cmidrule(r){4-5} 
%\cmidrule(r){6-7}
 parameter & \(x_i\) & \(\Delta x_i\) & \(x_i\) & \(\Delta x_i\) & \(x_i\) & \(\Delta x_i\)    \\
\mr
\(\rho_c \) & 0.161 & 0.004 & 0.159 & 0.004 & 0.156 & 0.004 \\
\(E_{sym}^{NM} / A\) & -16.056 & 0.02 &  &  &  &  \\
\(K_{sym}^{NM}\) &  &  &  &  & 239.930 & 2.0 \\
\(a_{sym}^{NM}\) & 30.543 & 0.1 & 28.987 & 0.2 & 29.131 &  0.2 \\ 
\(L_{sym}^{NM}\) & 45.080 & 0.4 & 40.005 & 0.4 &  & \\ 
\(1/M^*_s\) &  &  & 0.992 & 0.012 & 1.074 & 0.012 \\
\(C_{0}^{\rho \Delta \rho} \) & -55.261 & 0.6 & -45.135 & 0.6 & -46.831 & 0.6\\
\(C_{1}^{\rho \Delta \rho} \) & -55.623 & 2.0 & -145.382 & 2.0 & -113.164 & 2.0 \\
\(V_0^{n}\) & -170.374 & 2.0 & -186.065 & 2.0 & -208.889 & 2.0 \\
\(V_0^{p}\) & -199.202 & 2.0 & -206.580 & 2.0 & -230.330 &  2.0\\
\(C_{0}^{\rho \nabla J}\) & -79.531 & 0.7 & -74.026 & 0.7 & -64.308 & 0.7 \\
\(C_{1}^{\rho \nabla J}\) & 45.630 & 1.5 & -35.658 & 1.5 & -38.650 & 1.5 \\
\(C_{0}^{JJ}\) &  &  & && -54.433 & 2.0  \\
\(C_{1}^{JJ}\) &  &  & & & -65.903 & 4.0 \\
\br
\end{tabular}
}
\label{tab:parsteps}
\end{table}

%\clearpage

\end{document}